\documentclass{article}

\usepackage{arxiv}

\usepackage[utf8]{inputenc} % allow utf-8 input
\usepackage[T1]{fontenc}    % use 8-bit T1 fonts
\usepackage[numbers,sort&compress]{natbib}
\usepackage{hyperref}       % hyperlinks
\usepackage{url}            % simple URL typesetting
\usepackage{booktabs}       % professional-quality tables
\usepackage{amsfonts}       % blackboard math symbols
\usepackage{nicefrac}       % compact symbols for 1/2, etc.
\usepackage{microtype}      % microtypography
\usepackage{cleveref}       % smart cross-referencing
\usepackage{lipsum}         % Can be removed after putting your text content
\usepackage{graphicx}
\usepackage{doi}
\usepackage{here}
\usepackage{algorithm}
\usepackage{algorithmicx,algpseudocode}
\usepackage{makecell}
\title{ROFBS$\alpha$: Real‑Time Backup System Decoupled from ML‑Based Ransomware Detection}

\newif\ifuniqueAffiliation
\uniqueAffiliationtrue

\ifuniqueAffiliation % Standard variant of author block
\author{ \href{https://orcid.org/0009-0001-0594-4620}{\includegraphics[scale=0.06]{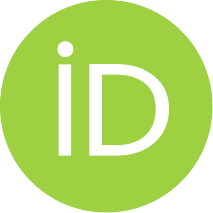}\hspace{1mm}Kosuke Higuchi} \\
	Kogakuin University\\ 1-24-2 Nishi-Shinjuku Shinjuku-ku\\ Tokyo, Japan\\
	\texttt{ed25001@ns.kogakuin.ac.jp} \\
	\And
	\href{https://orcid.org/0000-0001-5956-3455}{\includegraphics[scale=0.06]{orcid.pdf}\hspace{1mm}Ryotaro Kobayashi} \\
	Kogakuin University\\ 1-24-2 Nishi-Shinjuku Shinjuku-ku\\ Tokyo, Japan\\
	\texttt{ryo.kobayashi@cc.kogakuin.ac.jp} \\
}
\else
\usepackage{authblk}

\newbox{\orcid}\sbox{\orcid}{\includegraphics[scale=0.06]{orcid.pdf}} 
\author[1]{%
	\href{https://orcid.org/0000-0000-0000-0000}{\usebox{\orcid}\hspace{1mm}David S.~Hippocampus\thanks{\texttt{hippo@cs.cranberry-lemon.edu}}}%
}
\author[1,2]{%
	\href{https://orcid.org/0000-0000-0000-0000}{\usebox{\orcid}\hspace{1mm}Elias D.~Striatum\thanks{\texttt{stariate@ee.mount-sheikh.edu}}}%
}
\affil[1]{Department of Computer Science, Cranberry-Lemon University, Pittsburgh, PA 15213}
\affil[2]{Department of Electrical Engineering, Mount-Sheikh University, Santa Narimana, Levand}
\fi

\renewcommand{\shorttitle}{ROFBS$\alpha$: Independent Real‑Time Backup System}

\hypersetup{
pdftitle={ROFBS$\alpha$: Real‑Time Backup System Decoupled from ML‑Based Ransomware Detection},
pdfsubject={cs.CR},
pdfauthor={Kosuke Higuchi, Ryotaro Kobayashi},
pdfkeywords={defence model, backup system, file prevention, ransomware detection for machine learning, eBPF},
}

\begin{document}
\maketitle

\begin{abstract}
	This study proposes ROFBS$\alpha$, a novel defensive architecture designed to address the issue of detection time in machine learning-based ransomware detection models.
	Built upon our previously proposed Real-time Open-File Backup System (ROFBS), ROFBS$\alpha$ introduces an asynchronous design that separates the backup and detection processes.
	By leveraging eBPF to detect file access events and executing the backup operation as an independent process, the proposed system eliminates processing interference while enabling flexible and fast protection.
	Evaluation experiments using three types of ransomware—AvosLocker, Conti, and IceFire—were conducted, and the effectiveness of ROFBS$\alpha$ was assessed in terms of the number of encrypted and backed-up files, backup ratio, and detection time.
	The results demonstrate that ROFBS$\alpha$ is an effective defense architecture that achieves high backup success rates and fast detection while reducing system load and processing interference.
	However, further improvements are needed to defend against ransomware with extremely rapid encryption behavior.
\end{abstract}


\section{Introduction}
In recent years, ransomware-based cyberattacks have rapidly increased, causing serious damage not only to individual users but also to critical infrastructure such as healthcare facilities, government agencies, and energy systems \cite{Berrueta2022}.  
In particular, the emergence of Ransomware as a Service (RaaS) has enabled individuals without technical expertise to easily launch ransomware attacks, thereby expanding the range of potential attackers \cite{Raas2021}.  
As a result, both the frequency and the scale of ransomware attacks have increased each year, with reports indicating that 72.7\% of organizations worldwide were affected by such attacks in 2023 \cite{cyberedge}.  
Furthermore, as of 2022, the annual economic losses caused by ransomware were estimated at approximately \$265 billion, making it a pressing issue in the field of information security \cite{allianz}.  

In response to this situation, ransomware detection methods based on machine learning (ML) have attracted increasing attention in recent years.  
ML-based approaches offer flexible detection capabilities that do not rely on predefined patterns, and are considered effective against obfuscated malware and emerging variants \cite{Amjad2023}.  
According to Beaman et al., many studies have proposed ML models that classify ransomware based on process behavior, and it has been confirmed that, with sufficient training data, these models can achieve high detection accuracy \cite{Beaman2021}.  
In addition, the emergence of the extended Berkeley Packet Filter (eBPF) has made it possible to extract features for ML directly from kernel-level activities.  
Danyil et al. proposed a real-time ransomware detection system that integrates eBPF, ML, and natural language processing (NLP).  
Their system combines efficient data collection via eBPF, anomaly detection using ML, and text analysis through NLP, achieving a high detection accuracy of 94.7\% while significantly reducing the false positive rate \cite{Danyil2023}.  
Sekar et al. proposed a two-stage approach that combines eBPF and NLP to enable real-time detection and mitigation of ransomware.  
Their method successfully identified ransomware incidents within a few seconds of a zero-day attack with an impressive accuracy of 99.76\% \cite{Sekar2024}.  

However, one of the key challenges of such detection models is the existence of what is known as detection time—the delay between the initiation of ransomware activity and its actual detection \cite{Higuchi2025}.  
Since ML models require time for feature collection and classification, a certain amount of delay is unavoidable, during which ransomware may encrypt some files.  
Furthermore, as pointed out in the review by Kok et al., there is a lack of proposed defense models \cite{kok2019review}.  
In addition, there have been reports of cases in which backups themselves became targets of ransomware and were deleted or destroyed by attackers \cite{Maigida2019}.  
These issues indicate that the assumption 'successful detection equals damage prevention' does not necessarily hold true in practice.  

To address the issue of detection time inherent in conventional detection-centric approaches, we proposed a novel defense mechanism called the Real-time Open-File Backup System (ROFBS), which creates real-time backups immediately upon detecting a file open event \cite{Higuchi2025}.  
ROFBS leverages eBPF in a Linux environment to hook system calls with minimal overhead upon file access, enabling immediate backup of the target file and aiming to proactively prevent file encryption by ransomware.  

In this study, we propose ROFBS$\alpha$, a modified architecture based on the original ROFBS, in which the backup process is decoupled from the detection model.  
ROFBS$\alpha$ utilizes the output of the detection model as a trigger for the restoration process, while the defense mechanism itself operates as an independent, continuously running process.  
This configuration enables flexible file protection without being affected by detection time.  
We conduct evaluation experiments using three types of ransomware—AvosLocker, Conti, and IceFire—to quantitatively verify the effectiveness of the proposed method in terms of backup success rate.  

\section{Related work}
Hern\'andez et al. developed a system called “R-Locker,” which defends against ransomware by monitoring access to decoy (dummy) files \cite{gomez2018}.  
Their approach demonstrated that accessing a decoy file can serve as a trigger to terminate ransomware activity before encryption progresses.  

M. Medhat et al. proposed a hybrid approach that detects packed ransomware samples by scanning process memory dumps and dropped executables \cite{Mehnaz2018}.  
By enhancing the YARA rule framework to describe common ransomware characteristics, they achieved a detection rate of approximately 98\% for memory dump files.  

Zhuravchak et al. proposed a method that utilizes honeypot techniques to mitigate damage to the file system \cite{Zhuravchak2021}.  
This method identifies ransomware processes by monitoring access to honeypot files.  

Lee et al. noted that ransomware binaries often whitelist specific file extensions.  
They proposed a defense method that protects important files by randomly changing their extensions, successfully preventing encryption in 141 out of 143 ransomware samples \cite{Lee2019}.  

Kok et al. proposed a machine learning–based pre-encryption detection algorithm that can accurately detect ransomware before encryption occurs \cite{kok2019}.  
Their approach achieved a false positive rate of 1.56\% and an AUC (Area Under the Curve) of 99.3\%, demonstrating its potential as an effective preventive measure against various types of ransomware.  

Song et al. proposed an anomaly detection method for the Android platform based on resource metrics such as CPU usage, memory consumption, and I/O rates \cite{Song2016}.  
Their system detects and halts ransomware processes, prompting the user for confirmation before removing the malicious application, thereby protecting user data.  

Greg et al. built a stream processor on a programmable forwarding engine (PFE) and proposed a method for identifying malicious network behavior using a random forest classifier \cite{Greg2018}.  
Their approach demonstrated that ransomware could be detected before encryption by analyzing flow-based fingerprints.  

Fujinoki et al. proposed a backup system called PDPZR, which generates backups upon every data update and manages redundancy by removing outdated versions \cite{fujinoki2023}.  
Simulation results suggest that this method can effectively mitigate damage caused by zero-day ransomware.  
However, challenges remain regarding the configuration and tuning of automatic protection rules, as well as further validation in real-world environments.  

Gujar et al. proposed an automatic backup system using SSDs that detects newly added files in real time and immediately backs them up to a dedicated folder on the SSD \cite{Gujar2023}.  
Since the software resides entirely within the SSD and is isolated from the host machine, it offers stronger protection compared with conventional host-dependent software.  
However, challenges remain in optimizing the system for specific SSD models and reducing the processing overhead for real-time operations.  

Oujezsky et al. proposed the Intelligent Malware Defense System (IMDS), which uses AI and hash functions to prevent infections in backup data \cite{Oujezsky2023}.  
The system verifies file integrity before backup and blocks the process if anomalies are detected, thereby providing stronger protection than traditional passive backup methods.  

\section{Background knowledge}
\subsection{Detection Time}
ML-based detection models identify ransomware based on behavioral patterns and features extracted from training datasets.  
First, benign and malicious datasets are collected, and relevant features are extracted.  
Then, using public or custom datasets, the model is trained to classify new samples as either benign or malicious.  
The advantage of this approach is that it does not rely on predefined signatures or patterns, allowing it to detect new or variant ransomware strains.  
On the other hand, several challenges have been noted, including dependency on datasets and the existence of evasion techniques \cite{Amjad2023,Jung2017}.  

We consider detection time to be one of the main challenges in the current detection flow.  
Regardless of improvements in detection accuracy or speed, ML models inherently require a certain amount of time for feature collection and classification.  
The process of halting ransomware consists of three steps: feature collection, classification, and process termination.  
Here, we define the time required for feature collection and classification as response time, and the time between classification and process termination as kill time.  
The sum of these two constitutes the overall detection time.  

\begin{figure}[H]
	\begin{center}
	  \centerline{\includegraphics[width=0.6\columnwidth]{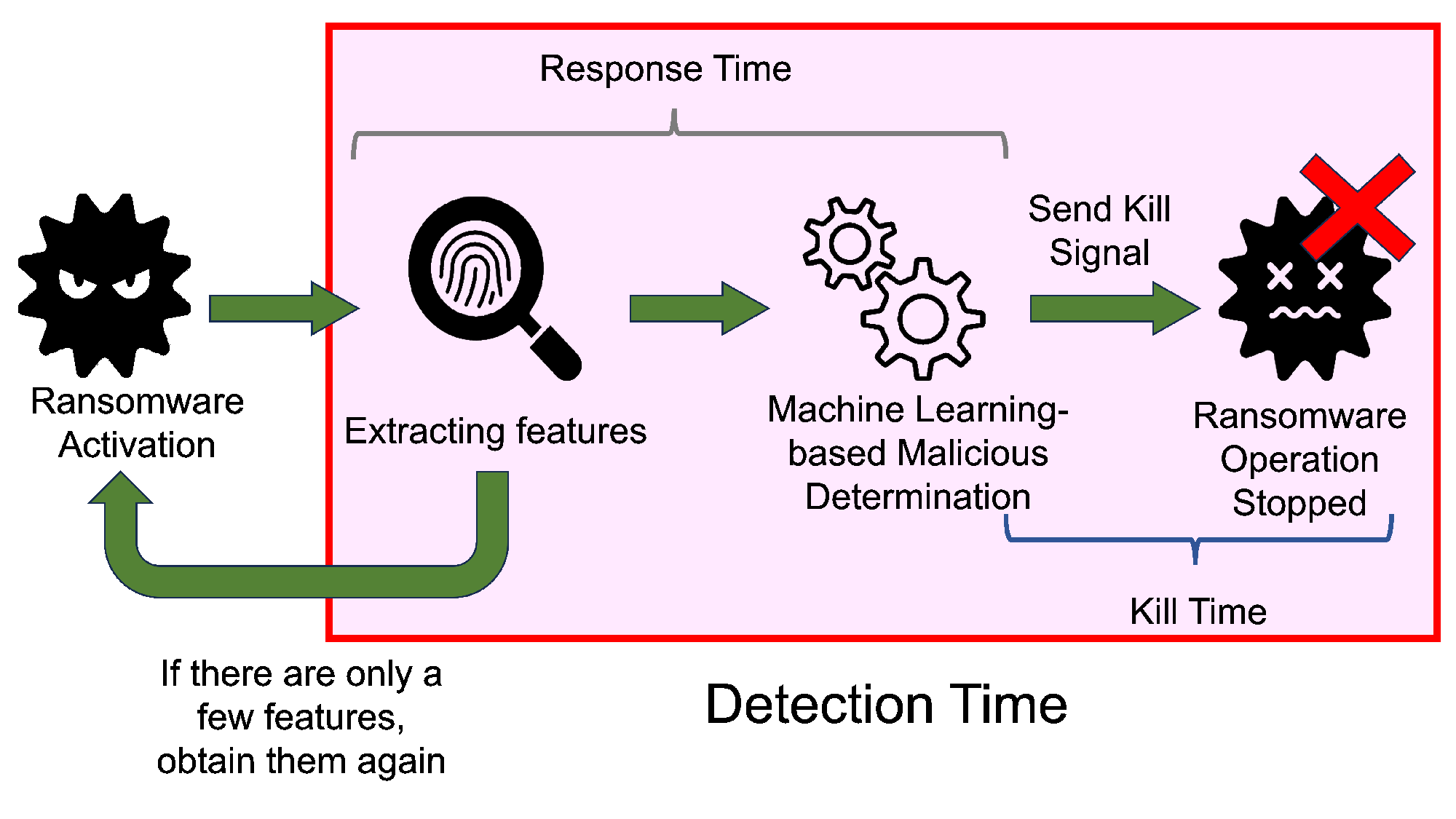}}
	  \caption{Breakdown of detection time into response time (feature collection and classification) and kill time (process termination).}
	  \label{fig:resvskill}
	\end{center}
  \end{figure}

Figure\ref{fig:resvskill} illustrates the relationship between response time and kill time in an ML-based detection model.  
Response time includes both feature collection and classification.  
While it can be shortened by improving feature extraction methods and model accuracy, it cannot be eliminated entirely.  

Although kill time tends to remain constant, it may increase under certain conditions, such as the presence of numerous child processes.  
Methods such as adjusting process priorities can be considered to minimize kill time.  
Reducing response time requires the adoption of more efficient feature extraction algorithms.  
By implementing these measures comprehensively, it is possible to reduce detection time and minimize the damage caused by ransomware.  
However, it is not possible to completely eliminate these time intervals.  

Therefore, it is important to employ defense mechanisms that function during detection time, rather than relying solely on the accuracy or speed of detection models.  

\subsection{extended Berkeley Packet Filter}
The extended Berkeley Packet Filter (eBPF) is a technology that allows user-defined programs to be executed safely and efficiently within a virtualized environment in the Linux kernel space \cite{eBPF}.  
Traditionally, extending kernel functionality required advanced tasks such as modifying the kernel source code or inserting kernel modules, which carried the risk of kernel crashes if not executed correctly.  
As a result, the introduction of eBPF has enabled safer and more flexible kernel extensions while avoiding such risks.  

eBPF enables flexible actions to be taken in response to events occurring within the kernel, and is widely used for various purposes such as application tracing, performance analysis, and security monitoring.  
Although eBPF functionality is generally dependent on the kernel version, a binary portability technique called Compile Once, Run Everywhere (CO-RE) has been developed to allow the same eBPF program to run across different environments.  
Furthermore, eBPF is designed to operate with minimal resource consumption, enabling stable performance with low memory usage while minimizing data loss.  

\subsection{BPF Compiler Collection}  
The BPF Compiler Collection (BCC) is a toolset for generating eBPF bytecode, developed by the IO Visor project \cite{bcc}.  
Although eBPF code that runs in kernel space must be written in C, Python or Lua can be used to load programs and process data, significantly improving development efficiency.  
In addition, BCC comes with a wide variety of utility tools, enabling developers to immediately use tracing and monitoring functionalities.  

eBPF uses a mechanism called probes to hook into kernel functions.  
There are two types: kprobe, which hooks at the function entry point, and kretprobe, which hooks after function execution.  
Since the information available before and after function execution differs, it is important to choose the appropriate type of probe depending on the intended purpose.  

\section{Proposal mechanism}
\begin{figure}[H]
  \begin{center}
    \includegraphics[width=0.7\columnwidth]{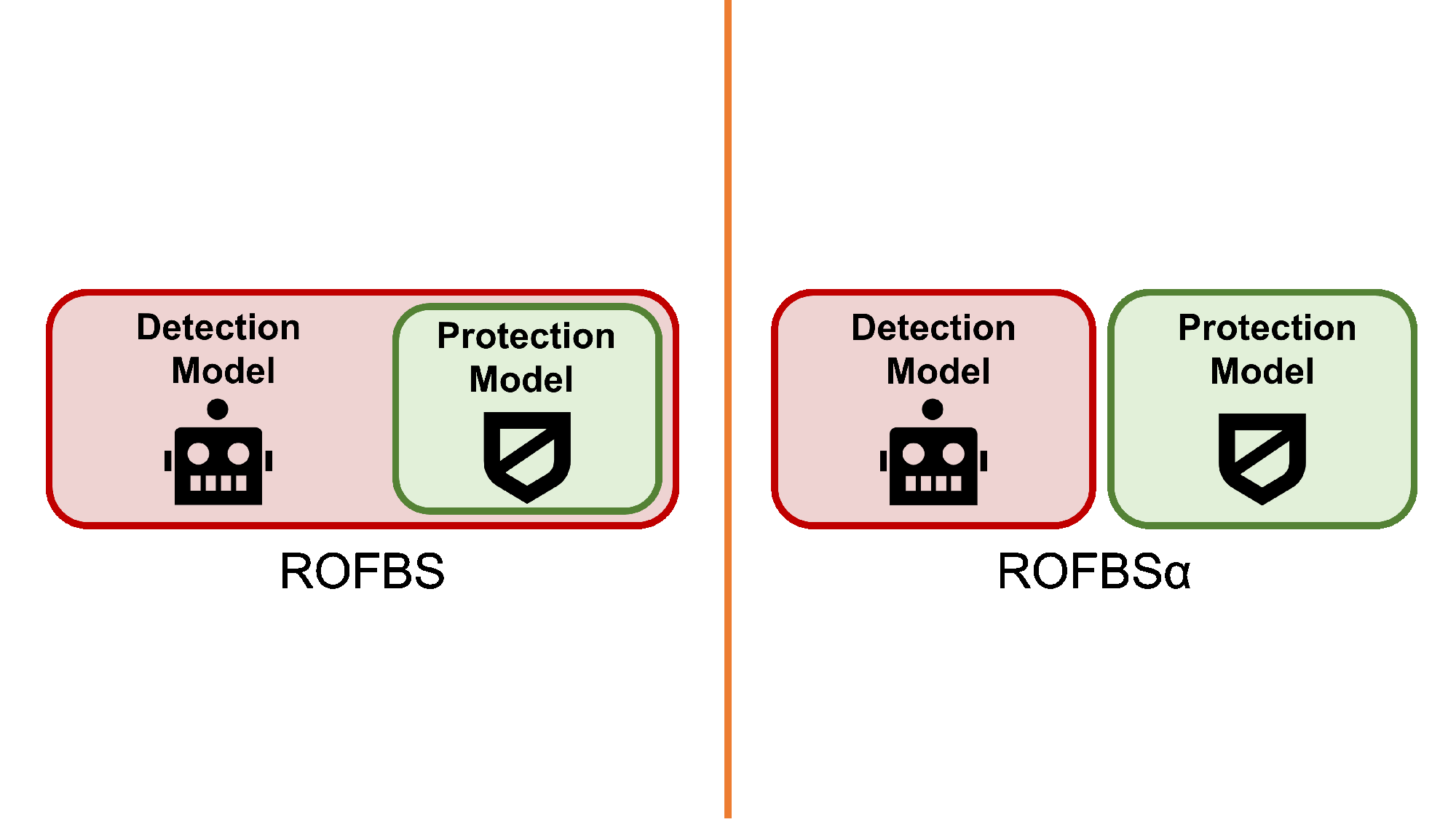}
    \caption{Comparison between ROFBS and ROFBS$\alpha$ architectures.}
    \label{fig:rofbsalpha}
  \end{center}
\end{figure}

Figure \ref{fig:rofbsalpha} illustrates the architectural differences between the conventional ROFBS and the proposed ROFBS$\alpha$.  

IIn the conventional ROFBS, the detection model and the defense mechanism are integrated into a single process, where detection and backup operations are performed sequentially.  
This implementation causes feature extraction to be blocked during the backup procedure, resulting in detection delays.  
In contrast, ROFBS$\alpha$ proposed in this study introduces an asynchronous architecture that completely separates the detection and defense mechanisms.  
Each component operates as an independent process, eliminating resource contention and allowing for non-blocking operation.  
This improvement provides the following advantages:  

\begin{enumerate}
  \item The detection and defense models do not compete for resources, thereby eliminating processing interference.  
  \item Feature extraction and inference by the detection model can continue during the backup process, significantly reducing detection time.  
  \item Parallel execution enhances both the backup success rate and detection speed, thereby improving overall system robustness and throughput.  
\end{enumerate}

As a result, ROFBS$\alpha$ improves the backup success rate while achieving faster detection without compromising the performance of the detection model.  
ROFBS$\alpha$ is triggered by the \texttt{xfs\_file\_open} kernel event in the Linux file system.  
When access to a target file is detected, its path and process information are passed to user space.  
Upon receiving this notification, ROFBS$\alpha$ sends a suspend signal to the corresponding process and immediately creates a backup file with the extension \texttt{.tmp} in the same directory.  
After confirming that the backup has completed, the process is resumed via a resume signal.  

File paths are received sequentially in reverse order from the kernel, and the system reconstructs the full path by combining multiple events.  
The use of the \texttt{.tmp} extension for backup files is based on prior findings that ransomware typically avoids encrypting files with certain extensions such as \texttt{.tmp} or \texttt{.exe} \cite{saleh2022}.  
Upon system termination, the backup files are renamed back to their original extensions.  

\makeatletter
\renewcommand{\ALG@name}{Algorithm: }
\makeatother
\begin{algorithm}[H]
  \caption{Backup Procedure of ROFBS$\alpha$}
  \label{algo:backup}
  \begin{algorithmic}[1]
    \Require File path ($file\_path$) and protected directories list ($protected\_directories$).  
    \Function{CreateBackupFile}{$file\_path$, $protected\_directories$}
      \If{$file\_path$ matches any directory in $protected\_directories$ \textbf{and} $file\_path$ has not yet been backed up}
        \State Create a temporary backup file named \texttt{$file\_path$.tmp}.  
        \State Mark $file\_path$ as backed up.  
      \EndIf
    \EndFunction
    \Function{RestoreIfMaliciousAndTerminated}{$file\_path$, $is\_malicious$}
      \If{$is\_malicious$}
        \State Rename \texttt{$file\_path$.tmp} to \texttt{$file\_path$}.  
      \EndIf
    \EndFunction
  \end{algorithmic}
\end{algorithm}

\section{Experiments}
\subsection{Experimental Setup}
We set up the experimental environment using VirtualBox.  
Table \ref{tab:evalution} summarizes the specifications of the host machine and the virtual machine.  

%Table
\begin{table}[H]
    \centering
    \caption{Experiment Environment}
    \scalebox{0.90}{
    \begin{tabular}{lcc}
        \hline
        & \textbf{Host Machine} & \textbf{Virtual Machine} \\
        \hline
        CPU     & Intel Core i7-12650H   & Virtual CPU (2 cores) \\
        Memory  & 32GB (DDR4 2400MT/s)   & 8GB                    \\
        Storage & 512GB (M.2 NVMe SSD)   & 40GB                   \\
        OS      & Windows 11             & AlmaLinux 9.1          \\
        \hline
    \end{tabular}}
    \label{tab:evalution}
\end{table}

The experiment was conducted based on prior research.  
We created a \texttt{victim} directory containing 4,385 files downloaded from the \texttt{tiny} folder of the NapierOne dataset \cite{Napierone2022}.  
The virtual environment includes all files and libraries required to run the detection model and ROFBS, along with the victim directory.  

We used three ransomware samples in this study, all of which were obtained from the internet between May and October 2023.  
The names of these ransomware samples follow the signature names given at download time.  

During execution, all default options were used except for the \texttt{--path} argument.  
Specifically, we used only arguments essential for execution or those listed in the help menu.  
For example, if the help menu listed a command-line argument \texttt{--path \$PATH}, we used that argument regardless of additional options such as thread counts or background execution.  
Ransomware often avoids encrypting certain extensions such as \texttt{.dll} and \texttt{.bat} to avoid crashing the operating system.  
Notably, IceFire does not support user-defined encryption targets.  

\begin{table}[H]
    \centering
    \caption{Ransomware Used for Experiments}
    \scalebox{0.90}{
    \begin{tabular}{lc}
        \hline
        \textbf{Ransomware} & \textbf{SHA256} \\
        \hline
        IceFire    & e9cc7fdfa3cf40ff9c3db0248a79f4817b170f2660aa2b2ed6c551eae1c38e0b \\
        AvosLocker & 0cd7b6ea8857ce827180342a1c955e79c3336a6cf2000244e5cfd4279c5fc1b6 \\
        Conti      & 95776f31cbcac08eb3f3e9235d07513a6d7a6bf9f1b7f3d400b2cf0afdb088a7 \\
        \hline
    \end{tabular}}
    \label{tab:ransomware}
\end{table}

The machine learning algorithms used in the experiment are listed in Table \ref{tab:ml_algo}.  
Accuracy and FPR were evaluated using logs obtained during the actual execution of the AvosLocker ransomware.  
These logs were unparsed and matched the format of those collected in this study.  
We manually labeled each log as benign or malicious.  
Accuracy was calculated based on agreement between the model's predictions and the manual labels.  
For example, if 2 out of 5 predictions matched the manual labels, the resulting accuracy would be $\frac{2}{5} = 0.4$, or 40\%.  
\begin{table}[H]
    \centering
    \caption{Machine Learning Algorithms Used}
    \scalebox{0.90}{
        \begin{tabular}{l r r}
            \hline
            \makecell[c]{\textbf{Algorithm}} & \textbf{Accuracy (\%)} & \textbf{FPR} \\
            \hline
            \textbf{Random Forest (RF)} & \textbf{97.2} & \textbf{0.0068} \\
            Gradient Boosting (GB)      & 96.8          & 0.0014         \\
            \hline
        \end{tabular}}
    \label{tab:ml_algo}
\end{table}

\subsection{Backup Ratio}
To evaluate the effectiveness of the defense model, we define the metric \textit{Backup Ratio} as follows:
\begin{equation}
  BackupRatio[\%] = \frac{Backup\_files}{Encrypted\_files} \times 100
\end{equation}

ROFBS and ROFBS$\alpha$ are proposed as countermeasures against the issue of detection time.  
Therefore, files that were not encrypted by ransomware are not relevant to the evaluation of these models.  
For this reason, we focus on the number of encrypted files that were successfully backed up.  
By examining the \textit{Backup Ratio}, we assess how effectively the proposed methods mitigate the damage caused by ransomware.  

Table \ref{tab:rofbs} shows the number of encrypted and backed-up files under the ROFBS model.  
Table \ref{tab:rofbsalpha} shows the corresponding results for ROFBS$\alpha$.  
In these tables, \textbf{E} denotes the number of encrypted files, and \textbf{B} denotes the number of backup files.  
Subscripts (e.g., $B_1$) indicate the results from the $N$-th execution of the experiment.  

\begin{table}[H]
    \centering
    \caption{Results for ROFBS: Encrypted and Backed Up Files}
    \scalebox{0.90}{
    \begin{tabular}{lcccccccc}
        \hline
        & \multicolumn{4}{c}{\textbf{RF}} & \multicolumn{4}{c}{\textbf{GB}} \\
        & $B_1$ & $E_1$ & $B_2$ & $E_2$ & $B_1$ & $E_1$ & $B_2$ & $E_2$ \\
        \hline
        AvosLocker & 29 & 29 & 35 & 35 & 116 & 131 & 85 & 121 \\
        Conti      & 32 & 32 & 36 & 36 & 34 & 34 & 34 & 34 \\
        IceFire    & 10 & 1 & 62 & 145 & 23 & 117 & 45 & 163 \\
        \hline
    \end{tabular}}
    \label{tab:rofbs}
\end{table}

\begin{table}[H]
    \centering
    \caption{Results for ROFBS$\alpha$: Encrypted and Backed Up Files}
    \scalebox{0.90}{
    \begin{tabular}{lcccccccc}
        \hline
        & \multicolumn{4}{c}{\textbf{RF}} & \multicolumn{4}{c}{\textbf{GB}} \\
        & $B_1$ & $E_1$ & $B_2$ & $E_2$ & $B_1$ & $E_1$ & $B_2$ & $E_2$ \\
        \hline
        AvosLocker & 0 & 0 & 2 & 2 & 43 & 43 & 19 & 19 \\
        Conti      & 13 & 13 & 18 & 18 & 36 & 36 & 36 & 36 \\
        IceFire    & 1 & 1 & 16 & 33 & 3 & 35 & 12 & 41 \\
        \hline
    \end{tabular}}
    \label{tab:rofbsalpha}
\end{table}

Next, Table \ref{tab:detectiontime_rofbs} presents the detection time for ROFBS.  
Table \ref{tab:detectiontime_rofbsalpha} shows the detection time for ROFBS$\alpha$.  
Here, \textit{first} indicates the detection time during the first execution.  
\textit{Second} refers to the detection time during the second execution.  

\begin{table}[H]
    \centering
	\caption{Number of Encrypted and Backed Up Files under ROFBS (per trial)}
		\scalebox{0.90}{
    \begin{tabular}{lcccc}
        \hline
        & \multicolumn{2}{c}{\textbf{RF}} & \multicolumn{2}{c}{\textbf{GB}} \\
        & First & Second & First & Second \\
        \hline
        AvosLocker & 0.18 & 0.47 & 0.16 & 0.17 \\
        Conti      & 0.28 & 0.29 & 0.32 & 0.32 \\
        IceFire    & 1.2  & 1.5  & 1.8  & 1.5 \\
        \hline
    \end{tabular}}
    \label{tab:detectiontime_rofbs}
\end{table}

\begin{table}[H]
    \centering
	\caption{Number of Encrypted and Backed Up Files under ROFBS$\alpha$ (per trial)}
    \scalebox{0.90}{
    \begin{tabular}{lcccc}
        \hline
        & \multicolumn{2}{c}{\textbf{RF}} & \multicolumn{2}{c}{\textbf{GB}} \\
        & First & Second & First & Second \\
        \hline
        AvosLocker & 0.08 & 0.10 & 0.16 & 0.14 \\
        Conti      & 0.18 & 0.22 & 0.44 & 0.45 \\
        IceFire    & 1.6  & 1.8  & 1.5  & 1.7 \\
        \hline
    \end{tabular}}
    \label{tab:detectiontime_rofbsalpha}
\end{table}

\section{Discussion}
Compared to the conventional ROFBS, ROFBS$\alpha$ significantly reduces the number of encrypted and backed‑up files across all ransomware attack scenarios.  
As shown in Tables \ref{tab:rofbs} and \ref{tab:rofbsalpha}, in the case of the AvosLocker attack using the RF-based detection model, both the number of encrypted files and the number of backed‑up files decreased from 29 (ROFBS) to 0 (ROFBS$\alpha$).  
This improvement can be attributed to the asynchronous architecture of ROFBS$\alpha$, which eliminates interference between the backup and detection processes.  

In particular, ROFBS$\alpha$ achieved a 100\% backup ratio in the AvosLocker and Conti attack scenarios, demonstrating protection performance equivalent to that of ROFBS.  
In the IceFire scenario with the RF-based model (second trial), the backup ratio improved from approximately 43\% (ROFBS) to about 48\% (ROFBS$\alpha$).  
However, with the GB-based detection model under the same IceFire attack, the backup ratio slightly decreased from around 20\% to 8.6\%.  
Despite this, ROFBS$\alpha$ successfully maintained or improved the backup ratio while reducing system overhead in most scenarios.  

Regarding detection time, the introduction of ROFBS$\alpha$ led to a reduction in most ransomware attack patterns when using the RF-based detection model, except for IceFire.  
As shown in Tables \ref{tab:detectiontime_rofbs} and \ref{tab:detectiontime_rofbsalpha}, for AvosLocker, the detection time decreased from 0.18 seconds to 0.08 seconds in the first trial and from 0.47 seconds to 0.10 seconds in the second trial—less than one-fourth of the original.  

In contrast, when using the GB-based detection model, detection time increased slightly in some scenarios.  
For example, under the Conti attack, the detection time increased from 0.32 seconds to 0.44 seconds (first trial) and from 0.32 seconds to 0.45 seconds (second trial).  
This is presumably due to the asynchronous nature of ROFBS$\alpha$, where prioritizing the backup process introduces overhead into the inference timing of the GB model.  

In summary, ROFBS$\alpha$ significantly reduces both the number of encrypted files and the number of backup operations compared to ROFBS, while maintaining or improving the backup ratio against major ransomware threats.  
Notably, in the AvosLocker scenario, the system completely suppressed both encryption and backup activity, showcasing the effectiveness of its asynchronous design.  
Additionally, ROFBS$\alpha$ reduced detection time when used with the RF model—by more than 75\% in the case of AvosLocker.  
Although slight increases in detection time were observed with the GB model, the backup ratio remained mostly stable or improved.  
Therefore, the increase in detection time is unlikely to be a critical issue in practical settings.  

Overall, ROFBS$\alpha$ successfully achieves a high level of both protection performance and processing efficiency by minimizing operational interference.  
However, further improvements may be necessary for ransomware with extremely fast encryption speeds.

\section{Conclusion}
In this study, we proposed ROFBS$\alpha$, a new architecture that separates the detection and defense models to reduce processing interference.  
Through evaluation experiments, we confirmed that ROFBS$\alpha$ significantly reduces both the number of encrypted files and the number of backed‑up files compared to the conventional ROFBS.  
Notably, in the case of an AvosLocker attack, the number of encrypted and backed‑up files was reduced to zero, which is attributed to the asynchronous architecture that eliminates interference between backup and detection processes.  

Regarding backup ratio, ROFBS$\alpha$ consistently maintained nearly 100\%, achieving protection performance comparable to or even exceeding that of ROFBS.  
Although a slight decrease in backup ratio was observed during the IceFire attack when using the GB‑based detection model, ROFBS$\alpha$ demonstrated a favorable trade‑off between protection and processing efficiency across major ransomware scenarios.  

Furthermore, regarding detection time, all attack patterns under the RF‑based detection model showed a significant reduction.  
In particular, the detection time for AvosLocker was improved to less than one‑fourth of the original value.  
While a slight increase in detection time was observed in some scenarios with the GB‑based model, this increase is attributed to inference timing interference resulting from asynchronous backup operations.  

Based on these results, ROFBS$\alpha$ is shown to be an effective defensive architecture that maintains high backup success rates and fast detection while reducing processing interference and system overhead.  
However, further improvements may be necessary to address ransomware with extremely fast encryption speeds, which remains a challenge for future work.  

\bibliographystyle{unsrt}
\bibliography{rofbsalpha}

\end{document}